\documentclass[preprint]{JHEP3} 

\usepackage{epsfig,multicol,bbm}

\newcommand\fverb{\setbox\fverbbox=\hbox\bgroup\verb}
\newcommand\fverbdo{\egroup\medskip\noindent%
			\fbox{\unhbox\fverbbox}\ }
\newcommand\fverbit{\egroup\item[\fbox{\unhbox\fverbbox}]}
\newbox\fverbbox

\title{A Note on a Certain Non-Gaussian Integral}

\author{Ulrik Svensson\\
	E-mail: \email{ulrik.svensson@bahnhof.se}}

\abstract{In this paper we present a general formula for the inhomogeneous non-Gaussian integral $I_d(S_1,S_2)=\int dx_1\ldots dx_d e^{-\frac{1}{2}S_1^2-S_2}$, where $S_1$ and $S_2$ are symmetric quadratic forms. The solution depends on the eigenvalues of the matrix $A=-iM_2M_1^{-1}$, where $M_1$ and $M_2$ are the matrix representations of $S_1$ and $S_2$ respectively. In the $2$-dimensional case we also give a manifestly $SO(2)$-invariant formulation in terms of invariants of the matrix $A$. An expression for $I(S_1,S_2)$ in the infinite-dimensional case is calculated and the solution depends only on the determinants of $M_1$ and $M_2$. The infinite-dimensional case may be of use in QFT.}

\keywords{Non-Gaussian integral, QFT}

\begin{document} 


\section{Introduction}
Non-Gaussian integrals with homogeneous symmetric forms of degree three and higher were recently studied in \cite{morozov} and several low-dimensional cases were solved. In this short note we investigate the non-Gaussian integral
\begin{equation}
I_d(S_1,S_2)=\int dx_1\ldots dx_d e^{-\frac{1}{2}S_1^2-S_2},
\end{equation}
where $S_1$ and $S_2$ are symmetric quadratic forms. This is the simplest case of a non-homogeneous, non-Gaussian integral. The problem is simplified by the fact that both $S_1$ and $S_2$ are quadratic since, as such they can be expressed in terms of matrices. We therefore rewrite the integral $I_d$ in the following form
\begin{eqnarray}
I_d(S_1,S_2)&=&\int dx_1\ldots dx_d e^{-\frac{1}{2}S_1^2-S_2}\nonumber\\
&=&\int dx_1\ldots dx_d e^{-\frac{1}{2}(x^T M_1 x)^2-x^T M_2 x}\nonumber\\
&=&\int dt \int dx e^{-t^2/2}e^{-x^T M_2 x+itx^T M_1 x}\nonumber\\
&=&\int dt \frac{e^{-t^2/2}}{\sqrt{det(M_2-itM_1)}}
\end{eqnarray}
where we have used the well-known result
\begin{equation}
\int  dx_1\ldots dx_d e^{-x^T M x}=\frac{1}{\sqrt{det{M}}}.
\end{equation}
In \cite{morozov}, A. Morozov and Sh. Shakirov calculate the integrals in terms of the $SL(d)$-invariants of the homogeneous symmetric form $S$. As it turns out that approach is not the most convenient for the present problem. Instead, we shall rewrite the integral to become a function of the eigenvalues of the matrix $A=-iM_2M_1^{-1}$, in the following way,
\begin{eqnarray}
I_d(S_1,S_2)&=&\int dt \frac{e^{-t^2/2}}{\sqrt{det(M_2-itM_1)}}\nonumber\\
&=&\frac{i^{d/2}}{\sqrt{det(M_1)}}\int dt \frac{e^{-t^2/2}}{\sqrt{det\left(t\mathbbm{1}-A\right)}}\nonumber\\
&=&\frac{i^{d/2}}{\sqrt{det(M_1)}}\int dt \frac{e^{-t^2/2}}{\sqrt{(t-z_1)\ldots (t-z_d)}}\nonumber\\
&\equiv& \frac{i^{d/2}}{\sqrt{det(M_1)}}J_d,
\end{eqnarray}
where $z_i$ is the $i$:th eigenvalue of the matrix $A$ and
\begin{equation}
J_d=\int dt \frac{e^{-t^2/2}}{\sqrt{(t-z_1)\ldots (t-z_d)}}.
\label{JD}
\end{equation}

\section{Ward identities}
In this section we investigate differential operators that annihilate $J_d$. Because of the high symmetry of the integrand they take a rather simple form. For simplicity, let us consider the one-dimensional case, where
\begin{equation}
J_1=\int dt \frac{e^{-t^2/2}}{\sqrt{t-z}}.
\label{endim}
\end{equation}
We would like to find an operator which turns the integrand into a full derivative. To that end, consider
\begin{eqnarray}
-z\frac{dJ_1}{dz}&=&\frac{1}{2}\int dt \frac{-ze^{-t^2/2}}{(t-z)^{3/2}}\nonumber\\
&=&\frac{1}{2}J_1-\frac{1}{2}\int dt \frac{te^{-t^2/2}}{(t-z)^{3/2}}\nonumber\\
&=&\frac{1}{2}J_1+\frac{3}{4}\int dt \frac{e^{-t^2/2}}{(t-z)^{5/2}}\nonumber\\
&=&\frac{1}{2}J_1+\frac{d^2 J_1}{dz^2},
\end{eqnarray}
which is equivalent to
\begin{equation}
\frac{d^2 J_1}{dz^2}+z\frac{dJ_1}{dz} +\frac{1}{2}J_1=0.
\label{endimdiff}
\end{equation}
We have so far not specified the contour of integration in equation \ref{endim}. It is crucial in the derivation of the Ward identity that the boundary terms vanish. To that end we only consider closed contours. In analogy with \cite{morozov} we denote an \emph{admissible} contour as a contour for which the integral \ref{endim} converges. In this case any contour that tends asymptotically to the lines $Arg(z)=0$ and $Arg(z)=\pi$ are admissible. To make the integrand single-valued we introduce a branch cut along the line $Arg(z)=0$ as indicated in figure \ref{myfigure}. Note that the contours shown in figure \ref{myfigure} are closed on the Riemann sphere if the infinitely remote point is taken into account.
\FIGURE{\epsfig{file=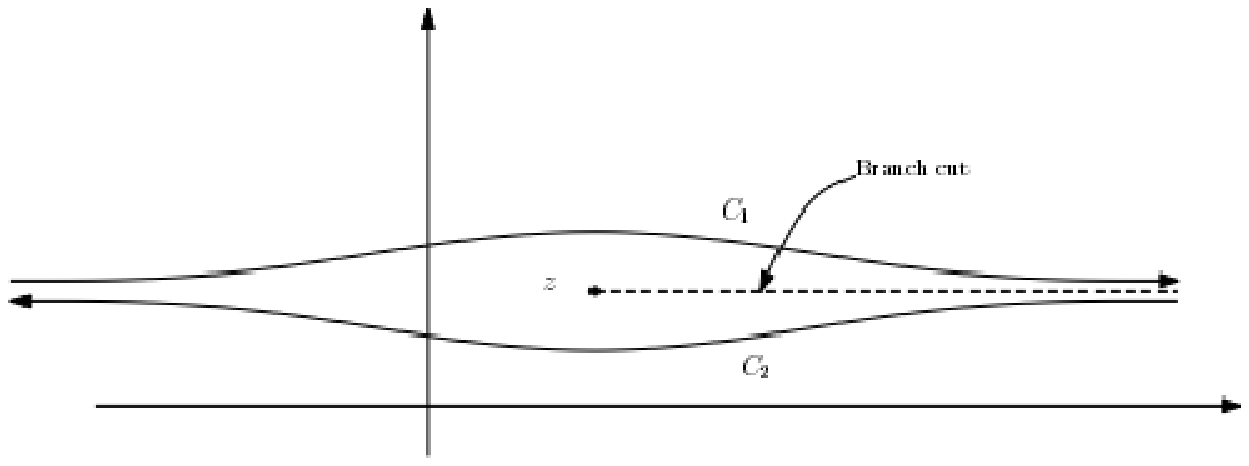,width=10cm} 
        \caption[Example of figure]{Contour of integration.}%
	\label{myfigure}}
There are two classes of admissible contours, and these classes are separated by the branch cut. The fact that there are two essentially different contours means that there will be two different solutions to the integral \ref{endim}. This can also be seen from that fact that the differential equation \ref{endimdiff} is second order. We denote the two independent solutions by $J_1^0$ and $J_1^1$.\\

An equally simple analysis for the $d$-dimensional case gives the following PDE:s:
\begin{equation}
\frac{\partial^2 J_d}{\partial z_k^2}+\sum_{i\neq k}\frac{\partial^2 J_d}{\partial z_k\partial z_i}+z_k\frac{\partial J_d}{\partial z_k}+\frac{1}{2}J_d=0, 
\label{diffeqn}
\end{equation}
where $k=1,\ldots,d$. In the $d$-dimensional case we have to distinguish between the cases when $d$ is even or odd. The explanation for this is that when $d$ is odd, the integrand of equation \ref{JD} has a branch point in the infinitely remote point, whereas in the even-dimensional case the eigenvalues $z_i$ are the only branch points. To see this, set $t=\frac 1z$ and $z_i=\frac 1\alpha_i$ in the denominator of \ref{JD}:
\begin{equation}
w=\frac{1}{\sqrt{(t-z_1)\ldots(t-z_d)}}=\frac{\sqrt{\alpha_1\ldots\alpha_d}z^{d/2}}{\sqrt{(z-\alpha_1)\ldots(z-\alpha_d)}}.
\label{branch}
\end{equation}
When $z$ is close to the origin, $w$ tends to
\begin{equation}
w=re^{in\theta/2},
\end{equation}
which shows that the origin is a branch point in the $z$-plane if and only if $d$ is odd. This in turn means that the infinitely remote point is a branch point for the integrand in equation \ref{JD} if and only if $d$ is odd.\\
\FIGURE{\epsfig{file=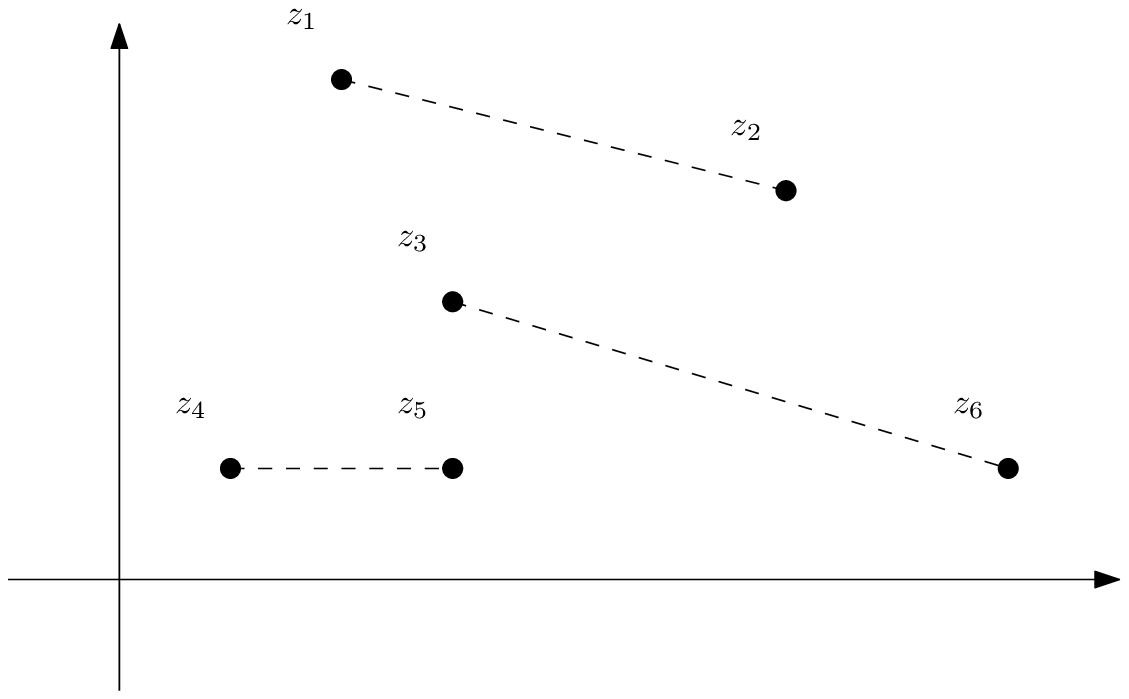,width=9cm} 
        \caption[Example of figure]{Branch cuts for the even-dimensional case.}%
	\label{even}}
In order to get a single-valued integrand, we must cut the plane to prevent contours to encircle the branch points. This is done by pairwise connecting the branch points with non-intersecting cuts. In the odd-dimensional case we also make a branch cut from one of the eigenvalues $z_i$ to the point at infinity. This is illustrated in figure \ref{even} and \ref{odd}.\\
\FIGURE{\epsfig{file=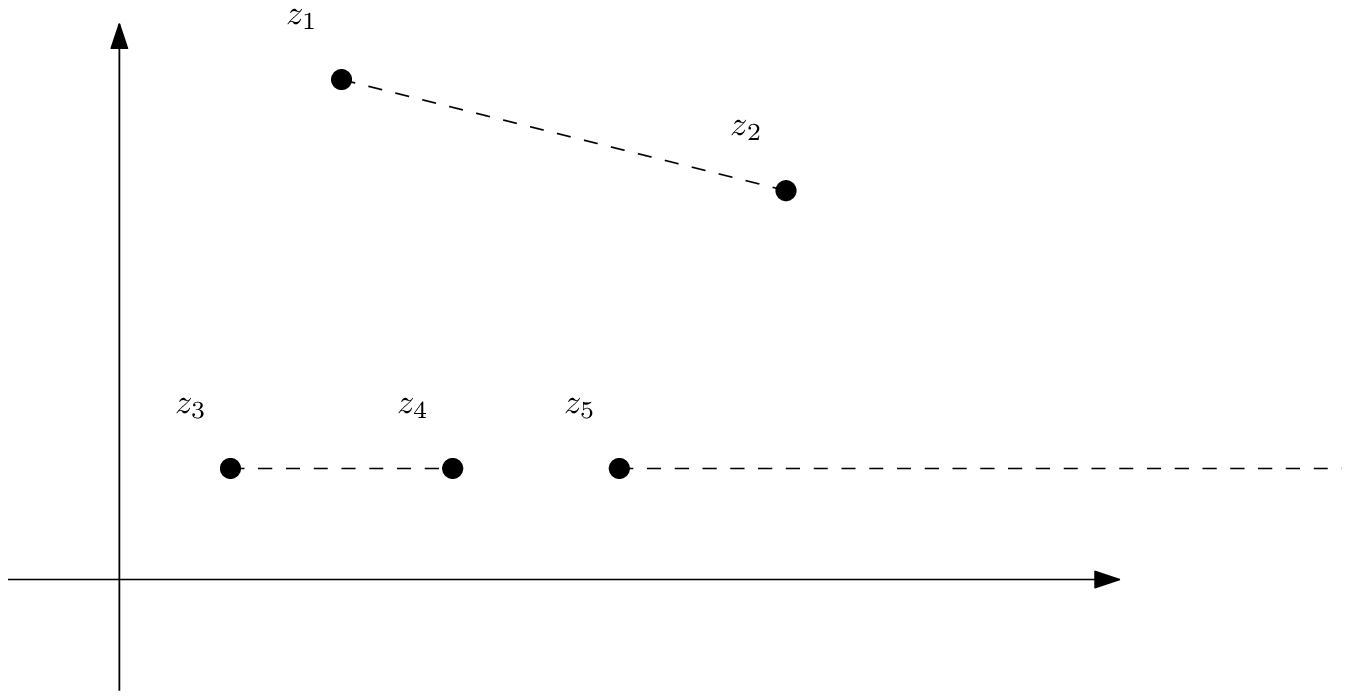,width=9cm} 
        \caption[Example of figure]{Branch cuts for the odd-dimensional case.}%
	\label{odd}}
In the even-dimensional case, any straight line not crossing any of the branch cuts is admissible if it is parallell to the real axis. In the odd-dimensional case, let $z_k$ be the branch point that is connected to the branch point at infinity. The admissble contours in this case are those that tend asymptotically to the lines $Arg(z_k)=0$ and $Arg(z_k)=\pi$ and does not cross any of the other branch cuts. In this way we get two different classes of contours for the odd-dimensional case. The two classes are separated by the branch cut from $z_k$ to the infinitely remote point. For the even-dimensional case there is only one class of admissble contours. The two independent solutions to equation \ref{diffeqn} are denoted $J_d^0$ and $J_d^1$, where $J_d^1$ is equal to zero in the even-dimensional case. 

\section{Solution of the Ward identities}

\subsection{Special cases}
Before we start with the general solution, we comment on some special cases. Consider the case when all the eigenvalues $z_i$ are equal to one. This means that $M_2=iM_1$. Moreover the integral
\begin{equation}
J_d(1,\ldots,1)=\int dt\frac{e^{-t^2/2}}{(t-1)^{d/2}}
\end{equation}
is just a constant so that 
\begin{equation}
I_d(S_1,S_2)=\frac{i^{d/2}}{\sqrt{det(M_1)}}J_d \propto \frac{1}{\sqrt{det(M_1)}},
\end{equation}
which is a direct consequence of the following equation that was proved in \cite{morozov}
\begin{equation}
\int e^{-S(x_1,\ldots,x_d)}dx_1\ldots dx_d\propto\int f\left( S(x_1,\ldots,x_d)\right)dx_1\ldots dx_d.
\end{equation}
Next, we consider the case when all the eigenvalues are equal, i.e. $z_1=\ldots=z_d=z$. In this case the differential equations \ref{diffeqn} reduce to the Kummer-equation,
\begin{equation}
\frac{d^2J_d}{dz^2}+z\frac{dJ_d}{dz}+\frac{d}{2}J_d=0,
\end{equation}
which has the solution
\begin{equation}
J_d(z,\ldots,z)=cU\left(\frac d4,\frac 12, \frac{z^2}{2}\right)+dM\left(\frac d4,\frac 12, \frac{z^2}{2}\right),
\end{equation}
where $U(a,b,z)$ and $M(a,b,z)$ are the Kummer $U$ and $M$ functions respectively and $c$ and $d$ are constants. Notice that in one dimension this leads to the following well-known result
\begin{equation}
\int ds \frac{e^{-t^2/2}}{\sqrt{t-z}}=\frac{1}{\sqrt{2}}\sqrt{z}e^{-z^2/4}K_{1/4}\left(\frac{z^2}{4}\right),
\label{bessel}
\end{equation}
where $K_{\nu}(z)$ is the modified Bessel-function of the second kind.

\subsection{All eigenvalues distinct and non-zero}
In this section we solve the differential equations \ref{diffeqn} when the eigenvalues of $A$ are distinct and non-zero. The solution is found using Frobenius method, i.e. by assuming the function $J_d$ can be expressed by an infinite series,
\begin{equation}
J_d=\sum_{i_1,\ldots,i_d =0}^{\infty}a_{i_1,\ldots,i_d}z_{1}^{i_1+k_1}\ldots z_{d}^{i_d+k_d},
\end{equation}
where the $k_i$ are some real numbers to be determined. Since the integral defining the function $J_d$ is completely symmetric in its arguments, it is clear that the coefficients $a_{i_1,\ldots,i_d}$ must be completely symmetric in their indices and that $k_1=k_2=\ldots =k_d$, leading to
\begin{equation}
J_d=\sum_{i_1,\ldots,i_d =0}^{\infty}a_{i_1,\ldots,i_d}z_{1}^{i_1+k}\ldots z_{d}^{i_d+k}.
\label{froebenius}
\end{equation}
Inserting this expression into equation \ref{diffeqn} gives the following algebraic equations
\begin{eqnarray}
a_{i_1+2,i_2,\ldots,i_d}&=&-\frac{i_2+k+1}{i_1+k+2}a_{i_1+1,i_2+1,i_3,\ldots,i_d}-\frac{i_3+k+1}{i_1+k+2}a_{i_1+1,i_2,i_3+1,i_4,\ldots,i_d}-\ldots \nonumber\\
&-&\frac{i_d+k+1}{i_1+k+2}a_{i_1+1,i_2,\ldots,i_d+1}-\frac{i_1+k+1/2}{(i_1+k+2)(i_1+k+1)}a_{i_1,\ldots,i_d},
\end{eqnarray}
and
\begin{equation}
k(k-1)=0.
\label{k-eq}
\end{equation}
Equation \ref{k-eq} is a consequence of the fact that any second order PDE will be a combination of two linearly independent functions. Note that in even dimension only $k=0$ yields a solution since the integral in this case is invariant under parity (i.e.\ the transformation $z_i\rightarrow -z_i$ for all $i$). For $k=0$ we get, after some algebraic manipulations, the following expression for the coefficients  
\begin{eqnarray}
a_{i_1,\ldots,i_d}&=&\frac{(-\frac 12)^{(i_1+\ldots+i_d)/2}\left(\frac 12\right)_{i_1}\ldots \left(\frac 12\right)_{i_d}}{i_1!\ldots i_d!\left(\frac d4 +\frac 12\right)_{(i_1+\ldots+i_d)/2}}A_0,
\label{coef-zero}
\end{eqnarray}
when $i_1+\ldots i_d=2n$ and zero otherwise. For $k=1$ they become
\begin{eqnarray}
a_{i_1,\ldots,i_d}&=&\frac{(-\frac 12)^{(i_1+\ldots+i_d)/2}\left(\frac 32\right)_{i_1}\ldots \left(\frac 32\right)_{i_d}}{(i_1+1)!\ldots (i_d+1)!\left(\frac{3d}{4} +\frac 12\right)_{(i_1+\ldots+i_d)/2}}B_0,
\label{coef-zero}
\end{eqnarray}
when $i_1+\ldots i_d=2n$ and zero otherwise. In these formul\ae, $(z)_{n}$ is the Pochhammer symbol defined by
\begin{equation}
(z)_{n}=z(z+1)(z+2)\ldots(z+(n-1)).
\end{equation}
In summary we have the following expressions for the two independent solutions to equation \ref{diffeqn}:
\begin{eqnarray}
J_d^{0}(z_1,\ldots,z_d)&=&A_0\sum_{i_1+\ldots+i_d=2n}\frac{(-\frac 12)^{n}\left(\frac 12\right)_{i_1}\ldots \left(\frac 12\right)_{i_d}}{i_1!\ldots i_d!\left(\frac d4 +\frac 12\right)_n}z_1^{i_1}\ldots z_d^{i_d},
\label{j-zero}
\end{eqnarray}
and 
\begin{eqnarray}
J_d^{0}(z_1,\ldots,z_d)&=&B_0\frac{det(M_2)}{det(M_1)}\times\nonumber\\
&&\times\sum_{i_1+\ldots+i_d=2n}\frac{(-\frac 12)^{n}\left(\frac 32\right)_{i_1}\ldots \left(\frac 32\right)_{i_d}}{(i_1+1)!\ldots (i_d+1)!\left(\frac{3d}{4} +\frac 12\right)_n}z_1^{i_1}\ldots z_d^{i_d},
\label{j-one}
\end{eqnarray}
where $J_d^{0}$ corresponds to $k=0$ and $J_d^{1}$ to $k=1$. To simplify the notation we set $B_0=0$ for $d$ even. The integral $I(S_1,S_2)$ is therefore given by
\begin{equation}
I_d(S_1,S_2)=\frac{1}{\sqrt{det(M_1)}}\left(J_d^0+J_d^1\right).
\label{solution}
\end{equation}
Taking the limit $d\rightarrow \infty$ only the terms corresponding to $n=0$ survive, leading us to 
\begin{equation}
I_{\infty}(S_1,S_2)=\frac{1}{\sqrt{det(M_1)}}\left(A_0+\frac{B_0det(M_2)}{det(M_1)}\right).
\label{solution-inf}
\end{equation}

\subsection{All eigenvalues doubly degenerate and $d$ even}
In this section we comment on the special case when $d$ is even and all the eigenvalues of $A$ are doubly degenerate. The differential equations in this case become
\begin{equation}
\frac{\partial^2 J_{DD}}{\partial z_k^2}+\sum_{i\neq k}\frac{\partial^2 J_{DD}}{\partial z_k\partial z_i}+z_k\frac{\partial J_{DD}}{\partial z_k}+J_{DD}=0, 
\label{diffeqn2}
\end{equation}
where the subscript $DD$ stands for double degeneracy and $k=1,\ldots \frac d2$. Notice that there are only $\frac d2$ distinct eigenvalues in this case. We solve these equations in the same way as we did in the previous section, and the result is
\begin{eqnarray}
J_{DD}^{0}(z_1,\ldots,z_{d/2})&=&A_0\sum_{i_1,\ldots,i_{d/2}=2n}\left(-\frac 12\right)^{n}\frac{1}{\left(\frac{\frac d2 +1}{2}\right)_n}z_1^{i_1}\ldots z_{d/2}^{i_{d/2}}.
\label{j-zero2}
\end{eqnarray}
Let us look att the first few terms in the sum \ref{j-zero2}. At the level $n=0$ there is just $A_0$. At level $n=1$, we have the following
\begin{equation}
-\frac{2A_0}{d +2}\left(z_1^2 + z_1z_2 + z_1z_3 +\ldots \right).
\end{equation}
In general at level $n$, we have
\begin{equation}
\left(-\frac 12\right)^{n}\frac{A_0}{\left(\frac{d+1}{2}\right)_n}H_{2n}(z_1,\ldots,z_{d/2}),
\end{equation}
where $H_n(z_1,\ldots,z_{d/2})$ is the complete symmetric polynomial of degree $n$ in the variables $z_i$. The expressions for the function $J^{0}_{DD}$ can therefore be written
\begin{eqnarray}
J_{DD}^{0}(z_1,\ldots,z_{d/2})&=&A_0\sum_{n}\left(-\frac 12\right)^{n}\frac{1}{\left(\frac{\frac d2 +1}{2}\right)_n}H_{2n}(z_1,\ldots, z_{d/2}),
\label{j-zero3}
\end{eqnarray}
leaving us with the following simple expression for the integral $I(S_1,S_2)$ in this case
\begin{eqnarray}
I_{DD}(S_1,S_2)&=\frac{A_0}{\sqrt{det(M_1)}}&\sum_{n}\left(-\frac 12\right)^{n}\frac{H_{2n}(z_1,\ldots, z_{d/2})}{\left(\frac{\frac d2 +1}{2}\right)_n}.
\end{eqnarray}

\subsection{$SO(d)$-invariance}
Even though rewriting the integral $I_d(S_1,S_2)$ in terms of the eigenvalues of $A=-iM_2M_1^{-1}$ greatly simplifies the solution, it has the imediate drawback of hiding the $SO(d)$-invariance of $I_d(S_1,S_2)$. To make this symmetry manifest we must replace the eigenvalues $z_i$ with invariants of $A=-iM_2M_1^{-1}$. In principle this can be done via the Cayley-Hamilton theorem which (in one form) states that
\begin{equation}
(t-z_1)\ldots(t-z_d)=t^d+I_1t^{d-1}+\ldots+I_{d-1}t+I_d,
\end{equation}
where $I_m$ are the invariants of $A=-iM_2M_1^{-1}$. Of course, this is an extremely difficult task except for the low-dimensional cases. In two dimensions we have
\begin{eqnarray}
z_1&=&\frac 12\left(I_1+\sqrt{I_1^2+4I_2}\right),\nonumber\\
z_2&=&\frac 12\left(I_1-\sqrt{I_1^2+4I_2}\right),
\end{eqnarray}
so that 
\begin{eqnarray}
J_2^{0}(I_1,I_2)&=&A_0\sum_{i,j=2n}\frac{(-\frac 12)^{n}\left(\frac 12\right)_i \left(\frac 12\right)_j}{i!j!n!}\times\nonumber\\
&& \times\left(\frac 12\left(I_1+\sqrt{I_1^2+4I_2}\right)\right)^{i}\left(\frac 12\left(I_1-\sqrt{I_1^2+4I_2}\right)\right)^{j}.
\end{eqnarray}
This expression gives the following manifestly $SO(2)$-invariant formulation of $I_2(S_1,S_2)$:
\begin{equation}
I_2(S_1,S_2)=\frac{1}{\sqrt{det(M_1)}}J_2^0(I_1,I_2).
\end{equation}

\section{Conclusion and Discussion}
In the present paper the non-Gaussian integral
\begin{equation}
I_d(S_1,S_2)=\int dx_1\ldots dx_d e^{-\frac{1}{2}S_1^2-S_2}
\end{equation}
has been studied. A general formula for any dimension is given in equation \ref{solution}. The formula is a function of the eigenvalues of the matrix $A=-iM_2M_1^{-1}$, where
\begin{eqnarray}
S_1&=&x^T M_1 x \mbox{ and}\nonumber\\
S_2&=&x^T M_2 x.
\end{eqnarray}
A formula for the infinite-dimensional case is given in equation \ref{solution-inf}, and depends only on the determinants of $M_1$ and $M_2$. This formula might be of use in quantum field theory. For two dimensions we give a formula with explicit $SO(2)$-invariance, replacing the eigenvalues with the invariants of $A$. The formulas given in equations \ref{j-zero}, \ref{j-one} and \ref{solution} might be of help to calculate the integral discriminant $J_{d|4}$, since any homogeneous action gives an inhomogeneous action when one of the coordinates is replaced by a constant, see equation (3) in \cite{morozov}. This has not been investigated in the present paper. \\

\end{document}